# Financial literacy, robo-advising, and the demand for human financial advice: Evidence from Italy

**David Aristei**
University of Perugia

**Manuela Gallo**
University of Perugia

**Abstract**

This paper investigates the impact of objective financial knowledge, confidence in one's financial skills, and digital financial literacy on individuals' decisions to seek financial advice from robo-advice platforms. Using microdata from the Bank of Italy's 2023 survey on Italian adults' financial literacy, we find that individuals with greater financial knowledge are less inclined to rely on online services for automated financial advice. Conversely, confidence in one's financial abilities and digital financial literacy enhance the likelihood of utilising robo-advice services. Trust in financial innovation, the use of digital financial services, and the propensity to take risks and save also emerge as significant predictors of an individual's use of robo-advice. We also provide evidence of a significant complementary relationship between the adoption of robo-advisory services and the demand for independent professional human advice. By contrast, a substitution effect is found for non-independent human advice. These findings highlight the importance of hybrid solutions in professional financial consulting, where robo-advisory services complement human financial advice.

**Keywords:** Robo-advisors; Financial literacy; Confidence; Digital financial literacy; Human financial advice; Fintech.

**JEL classification:** D14; G11; G41; G53; O33

# 1. Introduction

Technological innovation within the financial sector has significantly transformed the provision of financial advice, resulting in the rapid emergence of robo-advisors. These automated platforms, powered by algorithms and readily accessible online, offer increased efficiency and reduced costs for individual investors, ultimately leading to higher returns and improved portfolio performance (Mezzanotte, 2020; Rossi and Utkus, 2024).

In recent years, robo-advisory services have evolved from basic, non-personalised platforms to sophisticated systems offering tailored advice and fully automated portfolio management. First-generation models provided generic advice, suggesting potentially attractive investments, while subsequent models have integrated personalised recommendations with online execution, closely mirroring human advisory services (Maume, 2021).[1] As highlighted by Belanche et al. (2019), D'Acunto and Rossi (2023) and Reuter and Schoar (2024), this financial innovation has democratised access to sophisticated financial strategies previously available only to those who could afford human advisors. However, the lack of stable, empathetic human interactions and the feeling of making investment decisions alone can contribute to financial anxiety, especially for individuals with limited financial knowledge and digital skills (Caratelli et al., 2019; Alemanni et al., 2020).

Numerous studies have investigated the relationship between financial literacy and the likelihood of utilising financial advisory services. A part of this literature suggests that financial literacy and financial advice complement each other rather than substitute one for another (Calcagno and Monticone, 2015; Collins, 2012; Mazzoli et al., 2024). Moreover, empirical evidence suggests that individuals with higher levels of financial literacy are more likely to obtain accurate and higher-quality information from financial advisors (Bucher-Koenen and Koenen, 2010). In line with this perspective, van Rooij et al. (2011) find that individuals with lower financial literacy tend to rely more on informal sources rather than professional financial advice. Chauhan and Dey (2020) provide evidence that highly literate investors recognise the value of financial advice but are less likely to consult professional advisors if they demand relatively higher fees. Other authors, however, find no significant relationship between objective financial literacy and financial advice seeking (e.g., Kramer, 2016).

Recent literature has analysed the effect of financial literacy on individuals' adoption of robo-advisory services (for a review, see Cardillo and Chiappini, 2024; Nourallah et al., 2025).

---

[1] D'Acunto and Rossi (2021) proposed a taxonomy of robo-advisors that hinges on four defining dimensions: personalisation of advice, investor discretion, investor involvement, and human interaction.

However, empirical findings are still not conclusive. Isaia and Oggero (2022) show that individuals with higher objective financial literacy are more likely to be potential users of robo-advisors as they are better equipped to understand and navigate automated financial tools. Conversely, Fan and Chatterjee (2020) demonstrate that both basic and advanced objective financial knowledge are negatively associated with the probability of adopting robo-advisory services among US investors. Similar evidence is provided by Piehlmaier (2022), who shows that investors with greater financial knowledge are less likely to use robo-advice.

Individuals' confidence in their actual financial knowledge might also affect financial advice-seeking behaviour. Excessive confidence in one's own financial abilities is often associated with risky financial behaviours, as overconfident investors may underestimate the complexities of financial markets, resulting in negative outcomes (Pikulina et al., 2017; Aristei and Gallo, 2021). Previous studies have demonstrated that investors' tendency to overestimate their financial knowledge also plays a key role in affecting their willingness to seek financial advice (Guiso and Jappelli, 2006; Hackethal et al., 2012). In this regard, Kramer (2016) and Lewis (2018) demonstrate that unjustified confidence in one's own financial literacy significantly reduces individuals' propensity to seek professional human financial advice. Recent literature has also analysed the impact of self-assessed financial knowledge on investors' propensity to adopt robo-advice. In this respect, robo-advisors may enable investors to receive impartial expert advice in a comfortable setting without the embarrassment of interacting with a human advisor, thereby addressing the overconfidence bias in financial decision-making. Bai (2024) notes that limited subjective financial knowledge can hinder the effective use of robo-advisory services, resulting in a greater reliance on traditional human financial advisors. Similarly, Fan and Chatterjee (2020) demonstrate that subjective basic financial knowledge and subjective advanced investment knowledge are both positively associated with the propensity to adopt robo-advisors. Piehlmaier (2022) also points out that overconfident investors are characterised by a significantly higher likelihood of adopting robo-advice in a pre-chasm market, while increased financial literacy seems to be correlated with a lower tendency to use robo-advice. Individuals who feel more financially competent are more likely to embrace automated services, trusting the robo-advisors' capacity to allocate and manage their investments effectively. Conversely, those with lower perceived financial knowledge may hesitate to use these services, fearing that they will not properly understand how algorithmic advice and automated decision-making work.

The use of robo-advisory services is also influenced by individuals' digital skills and digital financial competencies, as online platforms for automated financial advice are inherently technology-driven and require users to have a certain ability to navigate them effectively. In this regard, Isaia and Oggero (2022) demonstrate that financial behaviours, such as online purchasing and digital payments, are positively associated with interest in receiving financial advice through digital platforms. In contrast, online activities without a financial component are not correlated with interest in robo-advisory services. Other studies examine the role of digital skills in the adoption of robo-advice platforms or focus on individuals' readiness to engage with new technologies. Flavián et al. (2022) report that feelings of insecurity regarding technology negatively affect users' intentions to adopt robo-advisors, while a certain level of discomfort with technology surprisingly has a positive influence on the intention to use such services. Filiz et al. (2022) demonstrate that even an individual's tendency towards algorithm aversion represents a significant obstacle to the broader adoption of robo-advisors.

Robo-advisory services constitute a viable alternative to traditional investment advisory models, particularly for investors concerned about potential conflicts of interest associated with human financial advisors or those who exhibit a high degree of confidence in regulatory mechanisms designed to protect against fraud. Specifically, Brenner and Meyll (2020) point out that robo-advisors are less susceptible to conflicts of interest, offering significantly lower and more transparent fee structures compared to human advisors, whose recommendations may be biased due to incentive-based compensation schemes. Nonetheless, users need to be confident that these platforms prioritise their needs and provide reliable advice (Bhatia et al., 2020; Nourallah et al., 2023). In this regard, Scherer and Lehner (2025) argue that trust in financial markets and financial service providers also plays a crucial role in the adoption of robo-advisory services, showing that investors who proactively verify their advisor's credentials are more likely to trust and use robo-advisors.

For investors with advanced financial knowledge, robo-advising represents an efficient tool for making investment decisions, effectively substituting human advice. Conversely, human advisors are essential for investors seeking personalised advice, especially in complex financial matters. In these cases, robo-advisors can complement human expertise by managing routine tasks, allowing human advisors to focus on more strategic and high-value activities; therefore, robo-advisors and human advisors should not necessarily be considered mutually exclusive. As pointed out by D'Acunto et al. (2019), hybrid advisory models, which combine the computational strengths of robo-advisors with the personalised insights of human advisors, are emerging as a promising solution. The most advanced robo-advisors autonomously manage portfolios, including rebalancing according to pre-defined investment strategies. Contemporary frameworks often adopt hybrid or blended models, combining algorithmic efficiency with human oversight (Maume, 2021).

In this paper, we contribute to the literature by examining the main factors affecting the adoption of robo-advisory services in Italy. First, we analyse the impact of objective financial literacy and confidence in one's financial competencies on the likelihood of using robo-advice. We further assess the role played by individuals' digital financial literacy, risk propensity, trust in technological and financial innovation, and use of digital financial products. Moreover, we explore the factors determining the intensity of robo-advice usage and investigate whether investors' adoption of robo-advisory services substitutes for or complements their demand for human financial advice. Only a few studies have examined the complementarity versus substitutability effects between robo-advisors and human advisors (Brenner and Meyll, 2020) or between individuals' financial competencies and various measures of human advice (Mazzoli et al., 2024). However, to the best of our knowledge, no studies have so far explored the relationships between robo-advising and different forms of dependent and independent human consulting. Additionally, our study contributes to the existing literature by examining the role of digital financial literacy in shaping individuals' adoption of digital platforms for financial advice. We therefore believe that our findings offer valuable insights with potential implications for both research and practice.

The remainder of the paper is organised as follows. Section 2 describes the data and defines the variables. In Section 3, we present and discuss the main empirical findings related to robo-advice usage and carry out several robustness checks. Section 4 examines the relationship between the use of robo-advice and the demand for professional human advice. Section 5 offers some concluding remarks.

## 2. Data and measurement

We use microdata from the latest release of the Bank of Italy's survey on adults' "Financial Literacy and Digital Financial Skills in Italy" (IACOFI), which was conducted in 2023 through Computer-Assisted Web Interviews on a nationally representative sample of about 5,000 individuals aged 18 to 79. The survey is based on the harmonised methodology developed by the OECD's International Network for Financial Education (OECD, 2022) and provides detailed information about respondents' financial behaviour, attitudes, and knowledge, as well as digital financial competencies, use of financial products, advice-seeking, and financial well-being.

This study examines the main factors influencing the use of robo-advisory services among Italian adults. To this aim, we consider as our main dependent variable a dummy (*Robo-advice*) equal to 1 if the respondent consulted an online platform for automated financial advice within the last 12 months. We also account for the intensity of robo-advice usage through an ordered variable indicating how frequently the respondent accessed platforms for financial robo-advising. As shown in Figure 1, approximately 70% of the respondents have never used robo-advisory services. In comparison, around 30% have sought robo-advice at least occasionally, and about 15% often and very often recur to robo-advisors. It is important to note that respondents who indicate having used a robo-advice platform may have solely sought advice and information, and may not have necessarily purchased financial products through the platform itself.

[Figure 1 about here]

Regarding the main explanatory variables, we exploit the seven test-based financial knowledge questions from the IACOFI questionnaire to define a measure of knowledge of basic financial concepts (i.e., time-value of money, interest paid on a loan, calculation of interest plus principal, compound interest rates, risk-return relationship, inflation, and risk diversification). Specifically, we define objective *Financial literacy* (FL) as the number of correct answers to the seven financial knowledge questions of the questionnaire. We then focus on individuals' confidence in their financial competencies through a measure of *Subjective Financial Literacy* (SFL), which ranges from 1 ("very low") to 5 ("very high"). Furthermore, we utilise the three test-based digital financial knowledge questions recently introduced in the IACOFI questionnaire, which relate to understanding digital financial contracts, personal data use, and crypto-assets, to construct a *Digital financial literacy* (DFL) score. This index is defined as the number of correct answers to the three digital financial knowledge questions of the questionnaire. Table 1 presents summary statistics for all the variables in the empirical analysis. On average, respondents correctly answer approximately 4 out of 7 questions on financial literacy and 1 out of 3 questions on digital financial literacy. At the same time, they assess their financial skills at a level of 2.4 on a scale from 1 to 5.

[Table 1 about here]

In our analysis, we control for individuals' objective *Financial well-being* through a score equal to the sum of four dummies (then normalised between 0 and 10), indicating whether respondents' income covers living expenses, if they can cover living expenses for at least three months in case of loss of their primary income source, if they can cope with a major expense without borrowing money, and whether they have money left at the end of the month. We include dummies indicating whether respondents saved money over the past 12 months (*Active saver*) and whether they are willing to risk some of their money when saving or making an investment (*Risk propense*). In line with Brenner and Meyll (2020), we also consider a binary variable equal to one for individuals who were victims of a financial scam within the last 24 months (*Financial fraud*). We then account for attitudes towards sustainable finance by using a dummy variable that identifies respondents who prefer investing in companies that minimise their environmental impact and strive to improve their social impact, risk management, ethics, and accountability (*Sustainable investor*). Moreover, we consider the intensity of respondents' use of digital payments and online money management over the past 12 months (*Digital payments*) through a score normalised to range from 0 to 10. Finally, as in Nourallah et al. (2023) and Scherer and Lehner (2025), we also assess the impact of individuals' trust in the financial services provided by online banks and FinTechs (*Trust in FinTech*) on the adoption of robo-advice.

To mitigate as much as possible omitted variable bias, we control for a large set of individual socio-demographic characteristics. Specifically, we include variables related to the respondent's gender and age (in years, included as both linear and quadratic terms), as well as binary indicators for educational attainment (*Upper secondary education* and *Tertiary education*) and working status (*Self-employed* and *Employee*). From Table 1, we observe that nearly 50% of the sample consists of women, with an average age of about 47 years. The respondents are predominantly employees (52.3%), and approximately 62% of them have at least an upper secondary education. We further control for a respondent's household disposable income (in classes), number of adults within the household, and include municipality size dummies (*Rural area/small municipality*) and macro-regional fixed effects.

Table A1 in the Appendix provides detailed definitions of the variables.

## 3. Results

*3.1 The determinants of robo-advice usage*

To assess the main determinants of individuals' demand for automated financial advisory services and shed light on the role of financial literacy and confidence, we consider the following probit model for the binary dependent variable of robo-advice usage $RA_i$:

$$RA_i = \mathbb{1}[\gamma FL_i + \delta SFL_i + \eta DFL_{it} + x_i'\boldsymbol{\beta} + \varepsilon_i > 0] \quad (1)$$

where $\mathbb{1}[\cdot]$ is an indicator function, $FL_i$ $SFL_i$ and $DFL_i$ represent objective, subjective and digital financial literacy, respectively, $x_i$ is a vector of exogenous regressors, and $\varepsilon_i$ is a standard normal error term.

Table 2 presents the marginal effects of regressors on the probability of robo-advisory usage, estimated from different probit models. In column 1, we first focus on the impact of objective financial literacy. The results reported reveal that higher financial knowledge has a negative and statistically significant (at the 1% level) impact on the likelihood of using robo-advisors, consistent with the findings of Fan and Chatterjee (2020). Specifically, a one-unit increase in the number of correct answers to basic financial knowledge questions reduces the probability of robo-advice usage by 3.41 percentage points. One plausible explanation for this finding is that individuals with high financial literacy may perceive the underlying mechanisms of advice, investment selection, and portfolio management by robo-advisors as insufficiently transparent and clear (Caratelli et al., 2019). Moreover, they might exhibit greater reluctance to disclose sensitive personal information through digital platforms (Scholz and Tertilt, 2021; Lee, 2009). At the same time, greater financial literacy may reduce the perceived discomfort typically associated with interactions with human financial advisors, thereby fostering a preference for human advisory services. Moreover, as suggested by Rossi and Utkus (2020), the propensity to use robo-advising is positively related to the need to acquire and expand knowledge on investment decisions, thus robo-advice could be considered a valuable tool to enhance financial competencies. It is worth noting that our findings differ from those of Isaia and Oggero (2022), who, however, focus on young adults in Italy and assess their intention to use robo-advisors, rather than their actual usage.

We then extend the empirical model to include subjective financial literacy as an additional control (column 2 of Table 2). This allows us to account for the role of what Parker and Stone (2014) label as *unjustified confidence*, that is, the residual effect of confidence in one's own financial abilities after controlling for objective knowledge. Although typically considered a detrimental cognitive bias, investor overconfidence may yield beneficial externalities, enhancing participation in financial markets (Xia et al., 2014) and increasing the propensity to adopt financial innovations (Piehlmaier, 2022). In this respect, our results provide evidence that individuals with greater unjustified confidence in their financial abilities are more inclined to seek advice from robo-advisors, supporting the significant role of overconfidence in fostering innovation in financial advice through robo-advisors. It is also noteworthy that the marginal effect of financial literacy remains negative and highly statistically significant. As discussed in Piehlmaier (2022), these findings suggest that ill-informed and overconfident investors play a critical role in the adoption and diffusion of an emerging financial service innovation characterised by a certain degree of information opacity and uncertainty, in line with the divergence of opinion (DOO) hypothesis (Miller, 1977).

Extending the model to account for the impact of digital financial literacy (column 3 of Table 2), we find that individuals with greater knowledge of digital financial concepts and products are significantly more inclined to use robo-advisors. Specifically, holding basic financial literacy constant, a one-unit increase in digital financial knowledge is associated with a 1.43 percentage point rise in the likelihood of using robo-advisory services. This result represents a novel contribution of our study to the literature. Existing research has, in fact, primarily focused on the role of algorithm aversion (Filiz et al., 2022), technology readiness (Flavián et al., 2022), and digital behaviours (Isaia and Oggero, 2022) in shaping individuals' intentions to use robo-advisory platforms. In this regard, our study is the first to provide empirical evidence suggesting that a lack of understanding of digital financial concepts represents a significant barrier to the adoption of robo-advisory services.

[Table 2 about here]

Regarding the other covariates, we find that respondents who exhibit greater financial well-being are less inclined to rely on robo-advisors. This evidence suggests that having difficulties in coping with unexpected income shocks without borrowing and covering living expenses using income or savings significantly increases an individual's propensity to resort to automated online platforms for financial advice. Furthermore, respondents who personally saved money during the last 12 months display a higher probability of resorting to robo-advisors to obtain support in managing their savings and investments. Accordingly, an individual's willingness to take financial risk strongly increases the adoption of robo-advisory services, consistent with the findings of Oehler et al. (2022) and Piehlmaier (2022). Our findings also suggest that respondents who use digital financial services, such as digital payments and online money management instruments, are significantly more likely to consult online platforms for automated advice. This finding aligns with the results of Isaia and Oggero (2022), who demonstrate that online activities with a financial component are predictive of robo-advice usage among young adults in Italy. Furthermore, in line with the results of Hildebrand and Bergner (2021), Flavian et al. (2022), and Scherer and Lehner (2025), we find that trust in online banks and FinTechs emerges as a key behavioural factor in robo-advice usage. This result emphasises that building trust among investors by enhancing transparency and reliability is crucial to fostering the adoption of robo-advisory services. We also find that having experienced a financial fraud during the last 24 months strongly increases robo-advice usage, in line with the findings of Bai (2021). Respondents who accepted advice to invest in a financial product that was later found to be a scam are 20.62 percentage points more likely to recur to robo-advisory services. Consistent with Brenner and Meyll (2020), who demonstrate that substitutability between robo and human advice is significantly driven by the fear of financial fraud, this result may indicate that investment fraud victimisation increases investors' distrust in human advisors and makes them more prone to rely on robo-advice. Interestingly, we further show that individuals concerned with the sustainability aspects of financial investments exhibit a greater willingness to utilise robo-advisory services. This finding provides valuable insights into the potential role of robo-advisors in encouraging sustainable investing among retail investors, as highlighted by Bruner and Laubach (2022) and Cardillo and Chiappini (2024)

Focusing on socio-demographic controls, results in column 3 of Table 2 indicate that women are 1.98 percentage points less likely to use automated platforms for financial advice, and the difference is significant at the 10% level. This evidence aligns with the findings of previous empirical studies (Rossi and Utkus, 2020; Seiler and Fanenbruck, 2021) and may be attributed to women being more risk-averse and less interested in financial investment and planning than men. Self-employed individuals are instead 4.91 percentage points more likely to use robo-advice than those not employed (i.e., unemployed, retired, students, homemakers, and those unfit for work). Conversely, age and educational attainment do not emerge as significant determinants of robo-advice usage. Consistent with Baulkaran and Jain (2023), respondents with a monthly disposable income higher than €1,750 are significantly more likely to recur to robo-advisory services. Finally, individuals residing in rural areas or small municipalities are significantly more likely to utilise online platforms for financial advice.

In column 4 of Table 2, we follow Piehlmaier (2022) and consider an alternative approach to assess the role of overconfidence on robo-advice usage. Specifically, we define a measure of unjustified confidence in one's own financial competencies as the residuals of a fourth-order polynomial regression of subjective financial literacy on financial and digital financial literacy.

The empirical findings remain substantially unchanged with respect to those reported in column 3 of the Table. Specifically, we find that a one-unit increase in overconfidence leads to a 1.43 percentage point rise in the probability of using a robo-advisor, confirming that excessive confidence in one's own financial abilities drives the adoption of robo-advice services.

We further extend our analysis and estimate an ordered probit model to assess the determinants of the intensity of robo-advice usage (Table 3). The empirical results reveal that a one-unit increase in financial literacy reduces the likelihood of utilising online platforms for financial advice sometimes, often, and very often by 0.94, 1.58, and 0.87 percentage points, respectively. Conversely, greater confidence in one's own ability to understand and apply basic financial concepts is positively associated with the probability of consulting robo-advisors. Specifically, a one-unit increase in respondents' subjective financial literacy corresponds to a 0.75 percentage point increase in the likelihood of consulting a robo-advisor at least occasionally. Furthermore, the probability of seeking online financial advice often or very often rises by 1.27 and 0.70 percentage points, respectively. The effect of digital financial literacy remains positive and statistically significant at the 1% level on the probability of accessing digital platforms for financial advice occasionally or frequently, while a one-unit increase in digital financial literacy results in a 0.25 percentage point rise in the likelihood of using robo-advisory services very often, with statistical significance at the 5% level. Finally, the influence of the other covariates remains essentially unchanged.

[Table 3 about here]

*3.2. Robustness*

In this Section, we carry out additional analyses to assess the robustness of our main findings.

We first examine the potential endogeneity of *Financial literacy*, *Subjective financial literacy*, and *Digital financial literacy* with respect to an individual's decision to use robo-advisory services. Prior studies have highlighted that endogeneity is a widespread issue that ought to be properly considered when evaluating the effect of financial literacy on financial outcomes (see, e.g., Lusardi and Mitchell, 2014; Stolper and Walter, 2017). In particular, financial and digital financial literacy might be endogenous determinants of robo-advice usage due to simultaneity and omitted variable bias. The advice received from automated online platforms may enhance users' financial knowledge, which could, in turn, affect their demand for financial advice. At the same time, some unobservable characteristics can drive individuals' demand for robo-advice and their willingness to acquire financial literacy. Piehlmaier (2022) also noted that prior use of robo-advice may inflate investors' confidence and result in an illusion of high financial knowledge, leading to potential endogeneity bias.

To address these concerns, we specify recursive systems of equations, consisting of a linear reduced-form equation for each potentially endogenous regressor and a probit equation for robo-advice usage. To identify the model parameters, the reduced-form equations must include at least one additional instrumental variable that induces exogenous variation in the potentially endogenous regressors and does not directly affect the outcome.

As additional instruments for financial literacy, we consider the average financial literacy that individuals of the same gender, residing in the same area, and belonging to the same age class as the respondent had in 2020 (derived from the 2020 release of the IACOFI survey) and a dummy indicating that the respondent is familiar with mortgage loans. Exposure to financial knowledge strongly influences individuals' level of financial literacy (van Rooij et al., 2011), while it does not directly impact the recourse to online platforms for automated financial advice. As discussed in Fornero and Monticone (2011), taking out a mortgage offers an opportunity for learning about basic financial concepts, which directly affects individuals' financial knowledge but not their use of robo-advice. Furthermore, we employ two additional instrumental variables for subjective financial literacy: a binary variable representing the respondent's propensity to purchase lottery tickets when perceiving a lack of financial resources, and a dummy variable indicating that the respondent believes their financial situation restricts their ability to pursue essential activities and exerts control over their life. The choice of these instrumental variables hinges on the idea that self-attribution, illusion of control, and optimism might strongly affect individuals' perception of their financial abilities, leading them to place unjustified confidence in their actual financial knowledge (Gervais and Odean, 2001; Williams and Gilovich, 2008; Kansal and Singh, 2018). Finally, the instrumental variables for digital financial literacy are the proportion of other individuals in the same area and age class as the respondent who reported being aware of crypto-assets and the previously discussed dummy regarding mortgage awareness. We expect that individuals' understanding of digital financial concepts will improve with their familiarity with mortgage loans and will be further enhanced as awareness of crypto assets becomes progressively widespread within their reference group.

Table 4 summarises the results from full maximum likelihood estimation of the three baseline probit models for robo-advice usage, extended to account for the potential endogeneity of *Financial literacy*, *Subjective financial literacy* and *Digital financial literacy*.[2] We first notice that the tests reported in the bottom part of the Table indicate that the overidentifying restrictions are valid and the additional instrumental variables are not weak, supporting our identification strategy. Furthermore, *Financial literacy* emerges as an endogenous determinant of robo-advice usage across all three specifications. Conversely, the exogeneity tests fail to reject the hypothesis of exogeneity of both *Subjective financial literacy* and *Digital financial literacy*. After accounting for potential endogeneity, the estimated marginal effect of financial literacy on the probability of robo-advice usage remains negative and significant at the 1% level. Still, it is larger in absolute terms than the one estimated from standard univariate probit models. Specifically, we find that a one-unit increase in the financial literacy score reduces the probability of using robo-advice services by about 11.29, 11.11, and 7.86 percentage points in the three specifications considered, respectively. This evidence provides strong empirical support for the significant substitutability between financial knowledge and the use of robo-advisory services. Additionally, it suggests that failing to account for endogeneity leads to downwardly biased estimates of the actual effect of financial literacy, as pointed out by Lusardi and Mitchell (2014). From Table 4, we also notice that the estimated marginal effects of subjective financial literacy and digital financial literacy remain positive and statistically significant and are broadly comparable to those reported in Table 2.[3]

[Table 4 about here]

We further evaluate the robustness of our main empirical findings to the definition of financial literacy. First, following van Rooij et al. (2011), we perform two separate factor analyses on the tetrachoric correlation matrices estimated on binary variables indicating the correct answer to each financial and digital financial knowledge question. We then use the predicted scores of the first factors to construct two continuous indexes of financial literacy and digital financial literacy, which we normalise to range between 0 and 7 and 0 and 3, respectively, to ease comparison with our baseline results. Second, as in Piehlmaier (2022), we estimate two-parameter Item Response Theory (2PL) models to assess a respondent's latent financial literacy and digital financial literacy. The estimated latent scores from these analyses (normalised between 0 and 7 and 0 and 3, respectively) are then used to define financial and digital financial literacy indexes. We also consider as an alternative measure of a respondent's financial literacy the number of correct answers to the Big Three questions, proposed by Lusardi and Mitchell (2008) and related to knowledge of interest rates, inflation, and risk diversification, which are commonly used in empirical analyses to assess an individual's level of basic financial knowledge. Finally, we extend our baseline measure of financial literacy to include an additional question related to understanding fixed-rate and adjustable-rate mortgage loans. The results of this robustness analysis are reported in Table 5 and confirm our baseline findings. The estimated marginal effects of financial literacy, subjective literacy, and digital financial literacy on the likelihood of using robo-advisory services are not influenced by the alternative definitions considered and remain largely consistent with those shown in Table 2.[4]

---

[2] Results remain largely unchanged when the endogenous probit models are estimated using a two-stage control function approach, as suggested by Wooldridge (2015). Control function estimates are available from the authors upon request.
[3] Similar results are obtained when we address potential endogeneity concerns in analysing how frequently individuals consult online platforms for automated financial advice. In particular, the results reported in Table A2 in the Appendix highlight that the estimated marginal effects of financial literacy on the probabilities of the different intensities of robo-advice usage have the same signs and statistical significance as those reported in Table 3, but they are larger in absolute terms. This evidence suggests that failing to account for endogeneity leads to an underestimation of the actual impact of financial literacy on the intensity of robo-advice usage.

[Table 5 about here]

## 4. Robo-advice usage and the demand for professional human advice

Focusing on the sub-sample of individuals who purchased at least one financial product within the last 24 months, we examine the relationship between the use of robo-advisory services and the demand for professional human advice. In this regard, the IACOFI survey provides detailed information on the relevance of alternative sources of financial advice, allowing us to distinguish between "independent" and "non-independent" professional human advisors based on their compensation structures and the extent to which external influences or incentives may impact their recommendations. Independent advisors are generally perceived as less biased and more aligned with the client's best interests, as they are not financially incentivised to sell specific products. In contrast, non-independent advisors may be subject to potential conflicts of interest due to their compensation models, which could compromise the objectivity of their advice. We thus extend the analysis of Brenner and Meyll (2020) and assess the demand for both types of professional human advice through the following bivariate probit model:

$$\begin{cases} IA_i = \mathbb{1}[\alpha_1 RA_i + \gamma_1 FL_i + \delta_1 SFL_i + \eta_1 DFL_i + x_i'\boldsymbol{\beta_1} + u_{1i} > 0] \\ NIA_i = \mathbb{1}[\alpha_2 RA_i + \gamma_2 FL_i + \delta_2 SFL_i + \eta_2 DFL_i + x_i'\boldsymbol{\beta_2} + u_{2i} > 0] \end{cases} \quad (2)$$

where $IA_i$ and $NIA_i$ are binary variables indicating the relevance of independent and non-independent human advice for an individual's decision to purchase financial products, respectively, and the errors $u_{1i}$ and $u_{2i}$ follow a standard bivariate normal distribution with arbitrary correlation $\rho$. The bivariate probit approach offers a wide range of options for interpreting estimation results, allowing us to estimate the marginal effects of regressors on both the marginal probabilities of the two binary outcomes and the joint probabilities of the different outcome combinations. Specifically, we can assess the impact of unit changes in regressors on the marginal probabilities of using independent and non-independent human advice (i.e., $P(IA = 1)$ and $P(NIA = 1)$) and on the joint probabilities of not using human advice ($P(IA = 0, NIA = 0)$), using independent advice only ($P(IA = 1, NIA = 0)$), using non-independent advice only ($P(IA = 0, NIA = 1)$), and using both types of advice ($P(IA = 1, NIA = 1)$).

---

[4] We also evaluate potential endogeneity concerns when employing these alternative definitions of financial literacy. The results reported in Table A3 confirm that financial literacy is an endogenous determinant of robo-advice usage, regardless of the measurement method considered. Furthermore, the average marginal effects of financial literacy remain negative and statistically significant at the 1% level and are larger in absolute terms than the corresponding effects reported in Table 5. This further confirms that neglecting to consider the endogeneity of financial literacy results in downwardly biased estimates of its effect on robo-advice usage.

In Table 6, we report the average marginal effects on the marginal probabilities of using independent and non-independent human advice. The empirical results indicate that users of robo-advisory services are 10.80 percentage points more likely to consult independent human advisors. Differently from Brenner and Meyll (2020), this evidence suggests a significant (at the 1% level) complementarity between the advice provided by automated online platforms and the demand for independent human advice. Robo-users are instead 5.81 percentage points less likely to seek advice from bank employees or the staff of financial services providers, suggesting that robo-advice substitutes for the potentially conflicted information provided by non-independent human advisors. From Table A4 in the Appendix, it is also noteworthy that the use of robo-advice significantly increases the likelihood of simultaneously seeking both types of human advice by 3.58 percentage points. This evidence further underscores the importance of hybrid solutions in professional financial consulting, where robo-advisory services complement recommendations and information provided by human financial advisors.

[Table 6 about here]

Our findings indicate that financial literacy significantly enhances the likelihood of seeking advice from non-independent advisors, whereas it only slightly increases the demand for independent human advice. Specifically, a one-unit increase in financial literacy is associated with a 3.09 percentage point rise (significant at the 1% level) in the probability of using non-independent advice and a 0.79 percentage point rise (only marginally significant at the 10% level) in the likelihood of consulting independent advisors. This complementary effect may suggest that individuals with higher financial literacy exhibit reduced concerns regarding their ability to recognise and address potential conflicts of interest that are inherent in non-independent human advice. These individuals, therefore, place greater importance on the value of human interaction in financial advising and leverage their enhanced financial knowledge to mitigate the potential for such conflicts of interest.

Digital finance literacy does not significantly impact individuals' demand for any form of professional human advice. From the analysis of the marginal effects on the joint probabilities (Table A4 in the Appendix), we notice that individuals with greater digital financial expertise are more inclined to rely solely on independent human advice and less likely to consult non-independent advisors exclusively, even though both effects are significant only at the 10% level.

In contrast to the findings of Mazzoli et al. (2024), individuals with higher levels of unfounded confidence in their financial abilities are more likely to regard the recommendations from independent financial consultants as important for their financial decisions. Conversely, self-assessed financial literacy does not significantly influence the demand for non-independent advisors. It is worth noticing that the indicators of human advice used in the analysis only refer to the relevance of the information and recommendations provided by advisors and do not imply the delegation of financial choices to external consultants. This may explain why individuals with greater financial confidence are more likely to prefer independent, fee-based human advisory services, which allow them to receive unbiased advice and make autonomous financial decisions.

We also find that the probability of resorting to independent fee-based financial advisors increases by 16.87 percentage points for individuals who accepted advice to invest in a financial product that was later found to be a scam. Recent experiences of financial fraud appear to lead potential investors to seek independent financial advice and distrust advisors who may have an interest in selling financial products issued or distributed by the financial institution they represent. In contrast, individuals who actively engage in saving tend to seek both types of professional human advice, whereas those who use digital payment systems and online money management tools exhibit a higher demand for non-independent advice only.

Focusing on socio-demographic characteristics, women are 6.58 percentage points less likely to seek non-independent advice than men, while no statistically significant difference is found for independent advisory services. Moreover, the likelihood of seeking advice from independent financial advisors significantly rises with the investor's age, while the probability of consulting non-independent advisors tends to decrease. Interestingly, individuals with a monthly disposable income lower than €1,751 are less inclined to consult independent advisors than those in higher income classes, whereas they display a significantly higher propensity to seek non-independent advice.

To deepen the analysis, we also evaluate the impact of the intensity of robo-advice usage on the demand for professional human advice. The average marginal effects reported in Table 7 indicate that as robo-advice usage intensifies, the complementarity between using online platforms for financial advice and the likelihood of seeking independent human consulting grows. Conversely, frequent or very frequent reliance on robo-advice decreases the likelihood of using the information provided by the staff of the financial product provider by 9.79 and 11.12 percentage points, respectively. In contrast, the occasional use of online platforms for automated financial advice does not affect the demand for non-independent human advice.

[Table 7 about here]

Additionally, we assess the robustness of our findings to potential sample selectivity issues that may arise from focusing on the subsample of individuals who have purchased financial products within the past two years. The results presented in Table A5 in the Appendix indicate that endogenous selection bias is not significant, confirming our baseline findings regarding the impact of robo-advisory usage on the demand for both types of professional human advice.

Following Brenner and Meyll (2020), we also address the potential endogeneity of robo-advice use with respect to the demand for professional human advice, which can arise due to simultaneity and omitted variable bias. To this end, we consider a recursive system of equations comprising the bivariate probit model in (2) and a reduced-form probit equation for the use of robo-advice. We use the dummy variable *Trust in FinTech* as an additional instrument for robo-advice usage, assuming that it exogenously affects the recourse to robo-advisory services, but it does not directly impact the demand for human advice. The results (columns 1 and 2 of Table A6 in the Appendix) suggest that the exogeneity of robo-advice usage cannot be rejected, providing evidence in favour of a causal relationship between the use of robo-advisory services and human advice-seeking. It is also worth noting that, after accounting for potential endogeneity, the estimated marginal effects of robo-advice usage on the demand for independent and non-independent human advice remain largely comparable to those reported in Table 6.

Finally, we address the potential endogeneity of financial literacy with respect to the propensity to seek professional human advice. To this end, we further extend the empirical model by adding a linear reduced-form equation for financial literacy, using the same instrumental variables discussed in Section 3.2. The results (columns 3 and 4 of Table A6 in the Appendix) confirm the exogeneity of robo-advice usage and, consistent with the findings of Calcagno and Monticone (2015) and Kramer (2016), suggest that financial literacy is an exogenous determinant of the demand for professional human advice. This evidence provides further support for a causal interpretation of our findings regarding the impact of robo-advice usage and financial literacy on the demand for both independent and non-independent human advice.

## 5. Concluding remarks

This paper examines the impact of financial literacy, confidence in one's financial competencies, and digital financial knowledge on the demand for financial advice provided by robo-advisors and human advisors. Specifically, exploiting novel data from the latest release of the Bank of Italy's IACOFI survey, our study first analyses the key determinants driving the adoption of robo-advisory services and the intensity of their use. Furthermore, we explore the relationship between the use of robo-advisors and the demand for different types of professional human advice. In doing so, we distinguish between "independent" and "non-independent" advisors, based on their compensation structures and the degree to which their recommendations may be shaped by external incentives or conflicts of interest.

The main empirical results suggest that individuals with higher levels of financial literacy are less likely to rely on online platforms for automated financial advice. In contrast, those who exhibit unjustified confidence in their financial abilities have a higher probability to engage with robo-advising. These findings underscore the pivotal role played by financially uninformed and overconfident investors in driving the adoption and diffusion of robo-advisory services. Moreover, we find that individuals with greater knowledge of digital financial concepts and products are significantly more likely to use robo-advisors. This evidence represents a novel contribution to the literature and highlights that limited understanding of digital finance represents a major barrier to the uptake of financial robo-advising. These findings are robust to several sensitivity checks, including controlling for potential endogeneity concerns. We also provide evidence of a significant complementary relationship between the use of robo-advisory services and the demand for independent human advice, which increases with the intensity of robo-advice usage. By contrast, we find that robo-advising substitutes the information provided by non-independent professional human advisors.

Overall, this study underscores that the successful development and widespread adoption of automated financial advisory systems cannot be achieved merely by improving individuals' basic financial literacy. Although individuals' financial knowledge is important for fostering a positive approach to financial advice and for achieving long-term financial well-being, as pointed out by an extensive body of literature, it is not sufficient to drive the adoption of robo-advice services. In this regard, targeted policy interventions aimed at enhancing digital financial competencies are essential. Such competencies encompass the ability to engage effectively with digital advisory platforms, an understanding of the features and risks associated with digital financial products and contracts, and the appropriate management and protection of personal data. Policy efforts must also address the psychological dimension of financial decision-making, particularly individuals' confidence in their own financial abilities, which significantly influences their willingness to engage with digital financial innovations. Furthermore, increased trust in financial intermediaries and fintech providers, along with strategic investments in transparency and system reliability, is likely to exert a significant positive influence on the adoption of robo-advisory services. Finally, our study emphasises the importance of developing hybrid financial advisory models, which combine the lower costs of robo-advice with the personalised support offered by human financial advisors, thereby improving accessibility for a broader range of investors.

# Tables

Table 1 – Summary statistics

| Variable | Mean | SD | Min | Max |
|---|---|---|---|---|
| Robo-advice | 0.298 | 0.457 | 0 | 1 |
| Independent human advice | 0.207 | 0.406 | 0 | 1 |
| Non-independent human advice | 0.391 | 0.488 | 0 | 1 |
| Financial literacy | 3.928 | 1.919 | 0 | 7 |
| Subjective financial literacy | 2.401 | 0.869 | 1 | 5 |
| Digital financial literacy | 1.278 | 0.999 | 0 | 3 |
| Financial well-being | 4.565 | 3.143 | 0 | 10 |
| Active saver | 0.816 | 0.387 | 0 | 1 |
| Risk propense | 0.094 | 0.292 | 0 | 1 |
| Trust in FinTech | 0.155 | 0.362 | 0 | 1 |
| Digital payments | 6.436 | 3.374 | 0 | 10 |
| Financial fraud | 0.042 | 0.202 | 0 | 1 |
| Sustainable investor | 0.349 | 0.477 | 0 | 1 |
| Female | 0.497 | 0.500 | 0 | 1 |
| Age | 46.893 | 15.893 | 18 | 79 |
| Number of adults | 2.226 | 0.963 | 1 | 7 |
| Self-employed | 0.092 | 0.289 | 0 | 1 |
| Employee | 0.523 | 0.500 | 0 | 1 |
| Upper secondary education | 0.414 | 0.493 | 0 | 1 |
| Tertiary education | 0.205 | 0.403 | 0 | 1 |
| Income: €1,751-€2,900 | 0.298 | 0.458 | 0 | 1 |
| Income: €2,900 or more | 0.077 | 0.267 | 0 | 1 |
| Rural area/small municipality | 0.400 | 0.490 | 0 | 1 |

**Notes**: The number of observations is equal to 4,391 for all the variables, with the exception of *Independent human advice* and *Dependent human advice* for which the number of observations is 1,249. The data are weighted and representative of the Italian population aged 18 and over.

Table 2 – The determinants of robo-advisory usage

| Model: | Probit | Probit | Probit | Probit |
|---|---|---|---|---|
| Dependent variable: | Robo-advice (1) | Robo-advice (2) | Robo-advice (3) | Robo-advice (4) |
| Financial literacy | -0.0341*** | -0.0355*** | -0.0369*** | |
|  | (0.0034) | (0.0034) | (0.0034) | |
| Subjective financial literacy | | 0.0314*** | 0.0297*** | |
|  | | (0.0080) | (0.0080) | |
| Digital financial literacy | | | 0.0143*** | |
|  | | | (0.0054) | |
| Overconfidence | | | | 0.0143*** |
|  | | | | (0.0036) |
| Financial well-being | -0.0154*** | -0.0160*** | -0.0160*** | -0.0203*** |
|  | (0.0019) | (0.0019) | (0.0019) | (0.0019) |
| Active saver | 0.0760*** | 0.0667*** | 0.0637*** | 0.0695*** |
|  | (0.0193) | (0.0188) | (0.0188) | (0.0195) |
| Risk propense | 0.0592*** | 0.0593*** | 0.0564*** | 0.0508*** |
|  | (0.0180) | (0.0175) | (0.0171) | (0.0170) |
| Trust in FinTech | 0.0646*** | 0.0654*** | 0.0649*** | 0.0693*** |
|  | (0.0115) | (0.0116) | (0.0118) | (0.0123) |
| Digital payments | 0.0853*** | 0.0834*** | 0.0828*** | 0.0860*** |
|  | (0.0026) | (0.0026) | (0.0027) | (0.0027) |
| Financial fraud | 0.2080*** | 0.2074*** | 0.2062*** | 0.2175*** |
|  | (0.0258) | (0.0252) | (0.0253) | (0.0270) |
| Sustainable investor | 0.0730*** | 0.0734*** | 0.0683*** | 0.0812*** |
|  | (0.0124) | (0.0124) | (0.0117) | (0.0124) |
| Female | -0.0241** | -0.0197* | -0.0198* | -0.0145 |
|  | (0.0117) | (0.0113) | (0.0113) | (0.0121) |
| Age | 0.0007 | 0.0006 | 0.0006 | 0.0004 |
|  | (0.0004) | (0.0005) | (0.0005) | (0.0005) |
| Number of adults | 0.0020 | 0.0007 | 0.0011 | 0.0036 |
|  | (0.0081) | (0.0080) | (0.0079) | (0.0082) |
| Self-employed | 0.0556** | 0.0508* | 0.0491* | 0.0435 |
|  | (0.0278) | (0.0287) | (0.0285) | (0.0300) |
| Employee | 0.0349* | 0.0311 | 0.0279 | 0.0275 |
|  | (0.0203) | (0.0208) | (0.0203) | (0.0210) |
| Upper secondary education | -0.0092 | -0.0173 | -0.0178 | -0.0293* |
|  | (0.0152) | (0.0147) | (0.0149) | (0.0152) |
| Tertiary education | 0.0721*** | 0.0715*** | 0.0716*** | 0.0551*** |
|  | (0.0105) | (0.0106) | (0.0105) | (0.0103) |
| Income: €1,751-€2,900 | 0.0925*** | 0.0952*** | 0.0933*** | 0.0779*** |
|  | (0.0213) | (0.0210) | (0.0213) | (0.0211) |
| Income: €2,900 or more | 0.0247** | 0.0238** | 0.0246** | 0.0272** |
|  | (0.0102) | (0.0103) | (0.0102) | (0.0110) |
| Rural area/small municipality | -0.0341*** | -0.0355*** | -0.0369*** | |
|  | (0.0034) | (0.0034) | (0.0034) | |
| McFadden pseudo $R^2$ | 0.3890 | 0.3924 | 0.3935 | 0.3728 |
| N | 4,391 | 4,391 | 4,391 | 4,391 |

**Notes**: The Table reports average marginal effects on the probability of using robo-advisory services, estimated from probit models. Robust standard errors, clustered by macro area, age class, and municipality size, are reported in parentheses below the estimates. All the models include macro area fixed effects.
\*\*\*, \*\* and \* denote significance at the 1, 5 and 10% levels, respectively.

Table 3 – The determinants of the intensity of robo-advisory usage

| Model: | Ordered Probit | | | |
|---|---|---|---|---|
| Dependent variable: | Robo-advice intensity | | | |
| Outcome: | Never | Sometimes | Often | Very often |
| | (1) | (2) | (3) | (4) |
| Financial literacy | 0.0339*** | -0.0094*** | -0.0158*** | -0.0087*** |
| | (0.0031) | (0.0009) | (0.0013) | (0.0012) |
| Subjective financial literacy | -0.0272*** | 0.0075*** | 0.0127*** | 0.0070*** |
| | (0.0070) | (0.0019) | (0.0031) | (0.0021) |
| Digital financial literacy | -0.0096* | 0.0026* | 0.0045* | 0.0025** |
| | (0.0050) | (0.0014) | (0.0024) | (0.0012) |
| Financial well-being | 0.0164*** | -0.0045*** | -0.0077*** | -0.0042*** |
| | (0.0016) | (0.0005) | (0.0008) | (0.0005) |
| Active saver | -0.0615*** | 0.0192*** | 0.0285*** | 0.0139*** |
| | (0.0158) | (0.0054) | (0.0071) | (0.0035) |
| Risk propense | -0.0639*** | 0.0159*** | 0.0300*** | 0.0181*** |
| | (0.0120) | (0.0029) | (0.0055) | (0.0041) |
| Trust in FinTech | -0.0339*** | 0.0089*** | 0.0159*** | 0.0091*** |
| | (0.0118) | (0.0031) | (0.0054) | (0.0034) |
| Digital payments | -0.0814*** | 0.0225*** | 0.0380*** | 0.0209*** |
| | (0.0026) | (0.0011) | (0.0020) | (0.0016) |
| Financial fraud | -0.1439*** | 0.0282*** | 0.0669*** | 0.0488*** |
| | (0.0146) | (0.0023) | (0.0071) | (0.0071) |
| Sustainable investor | -0.0716*** | 0.0194*** | 0.0337*** | 0.0184*** |
| | (0.0104) | (0.0029) | (0.0051) | (0.0030) |
| Female | 0.0199* | -0.0055* | -0.0093* | -0.0051** |
| | (0.0103) | (0.0030) | (0.0048) | (0.0025) |
| Age | -0.0007 | 0.0002 | 0.0003 | 0.0002* |
| | (0.0005) | (0.0001) | (0.0002) | (0.0001) |
| Number of adults | -0.0017 | 0.0005 | 0.0008 | 0.0004 |
| | (0.0078) | (0.0022) | (0.0037) | (0.0020) |
| Self-employed | -0.0208 | 0.0055 | 0.0097 | 0.0056 |
| | (0.0250) | (0.0064) | (0.0117) | (0.0069) |
| Employee | -0.0047 | 0.0013 | 0.0022 | 0.0012 |
| | (0.0181) | (0.0051) | (0.0085) | (0.0046) |
| Upper secondary education | -0.0009 | 0.0002 | 0.0004 | 0.0002 |
| | (0.0145) | (0.0040) | (0.0068) | (0.0037) |
| Tertiary education | -0.0522*** | 0.0141*** | 0.0245*** | 0.0136*** |
| | (0.0105) | (0.0028) | (0.0049) | (0.0030) |
| Income: €1,751-€2,900 | -0.0745*** | 0.0179*** | 0.0346*** | 0.0219*** |
| | (0.0181) | (0.0036) | (0.0081) | (0.0067) |
| Income: €2,900 or more | -0.0182* | 0.0050* | 0.0085* | 0.0047* |
| | (0.0103) | (0.0028) | (0.0049) | (0.0026) |
| Rural area/small municipality | 0.0339*** | -0.0094*** | -0.0158*** | -0.0087*** |
| | (0.0031) | (0.0009) | (0.0013) | (0.0012) |
| McFadden pseudo $R^2$ | 0.2596 | | | |
| N | 4,391 | | | |

**Notes**: The Table reports average marginal effects on the probabilities of the different intensities of robo-advice usage, estimated from an ordered probit model. Robust standard errors, clustered by macro area, age class, and municipality size, are reported in parentheses below the estimates. All the models include macro area fixed effects.
\*\*\*, \*\* and \* denote significance at the 1, 5 and 10% levels, respectively.

Table 4 – Robo-advisory usage: addressing potential endogeneity bias

| Model:<br>Dependent variable: | Endogenous probit<br>Robo-advice<br>(1) | Endogenous probit<br>Robo-advice<br>(2) | Endogenous probit<br>Robo-advice<br>(3) |
|---|---|---|---|
| Financial literacy | -0.1129***<br>(0.0105) | -0.1111***<br>(0.0108) | -0.0786***<br>(0.0062) |
| Subjective financial literacy |  | 0.0378***<br>(0.0068) | 0.0126***<br>(0.0026) |
| Digital financial literacy |  |  | 0.0082***<br>(0.0013) |
| Other control variables | Yes | Yes | Yes |
| Overidentification test | [0.3140] | [0.2880] | [0.4758] |
| Weak-instruments test: |  |  |  |
| *Financial literacy* | [0.0000] | [0.0000] | [0.0000] |
| *Subjective financial literacy* |  | [0.0000] | [0.0000] |
| *Digital financial literacy* |  |  | [0.0000] |
| Exogeneity test: |  |  |  |
| *Financial literacy* | [0.0000] | [0.0000] | [0.0000] |
| *Subjective financial literacy* |  | [0.2460] | [0.2690] |
| *Digital financial literacy* |  |  | [0.7650] |
| N | 4,391 | 4,391 | 4,391 |

**Notes**: The Table reports average marginal effects on the probability of using robo-advisory services, estimated from probit models extended to account for the potential endogeneity of financial literacy, subjective financial literacy, and digital financial literacy. Robust standard errors, clustered by macro area, age class, and municipality size, are reported in parentheses below the estimates. All the models include the same control variables as in Table 2 and macro area fixed effects. Complete results are available upon request. The additional instrumental variables for financial literacy are the average financial literacy that individuals of the same gender, living in the same area and belonging to the same age class as the respondent had in 2020 (derived from the 2020 Bank of Italy's IACOFI survey) and a dummy indicating that the respondent is familiar with mortgage loans. We use as additional instruments for subjective financial literacy a binary variable representing the respondent's inclination to purchase a lottery ticket when perceiving a lack of financial resources and a dummy indicating that the respondent believes their financial situation restricts their ability to pursue important activities and exerts control over their life. The instrumental variables for digital financial literacy are the proportion of other individuals in the same area and age class as the respondent who reported being aware of crypto-assets and a dummy indicating that the respondent is familiar with mortgage loans. The p-values of the overidentification, weak instruments, and exogeneity tests are reported in square brackets.

***, ** and * denote significance at the 1, 5 and 10% levels, respectively

Table 5 – Robo-advisory usage: alternative definitions of financial literacy

| Model: | Probit | Probit | Probit | Probit |
|---|---|---|---|---|
| Dependent variable: | Robo-advice | Robo-advice | Robo-advice | Robo-advice |
| Definition of FL: | IRT | Factor | Big Three | Eight questions |
|  | (1) | (2) | (3) | (4) |
| Financial literacy | -0.0410*** | -0.0367*** | -0.0499*** | -0.0330*** |
|  | (0.0034) | (0.0031) | (0.0057) | (0.0032) |
| Subjective financial literacy | 0.0303*** | 0.0289*** | 0.0285*** | 0.0313*** |
|  | (0.0080) | (0.0078) | (0.0079) | (0.0081) |
| Digital financial literacy | 0.0138*** | 0.0202*** | 0.0096* | 0.0145*** |
|  | (0.0053) | (0.0046) | (0.0055) | (0.0054) |
| Other control variables | Yes | Yes | Yes | Yes |
| McFadden pseudo $R^2$ | 0.4030 | 0.4042 | 0.3883 | 0.3948 |
| N | 4,391 | 4,391 | 4,391 | 4,391 |

Notes: The Table reports average marginal effects on the probability of using robo-advisory services, estimated from probit models considering alternative definitions of financial literacy. Robust standard errors, clustered by macro area, age class, and municipality size, are reported in parentheses below the estimates. All the models include the same control variables as in Table 2 and macro area fixed effects. Complete results are available upon request.
***, ** and * denote significance at the 1, 5 and 10% levels, respectively

Table 6 – Robo-advice usage and the demand for professional human advice

| Model: | Bivariate probit | |
|---|---|---|
| Outcome: | Marginal probabilities | |
| | Independent human advice | Non-independent human advice |
| | (1) | (2) |
| Robo-advice | 0.1080*** | -0.0581** |
| | (0.0323) | (0.0286) |
| Financial literacy | 0.0079* | 0.0309*** |
| | (0.0048) | (0.0078) |
| Subjective financial literacy | 0.0311** | 0.0196 |
| | (0.0135) | (0.0171) |
| Digital financial literacy | 0.0188 | -0.0187 |
| | (0.0132) | (0.0171) |
| Financial well-being | 0.0041 | 0.0106** |
| | (0.0044) | (0.0044) |
| Active saver | 0.0697*** | 0.0830** |
| | (0.0264) | (0.0390) |
| Risk propense | 0.0690** | -0.0331 |
| | (0.0348) | (0.0420) |
| Digital payments | 0.0040 | 0.0239*** |
| | (0.0044) | (0.0063) |
| Financial fraud | 0.1687*** | 0.0429 |
| | (0.0565) | (0.0763) |
| Sustainable investor | -0.0334 | -0.0106 |
| | (0.0213) | (0.0286) |
| Female | -0.0056 | -0.0658*** |
| | (0.0169) | (0.0248) |
| Age | 0.0032*** | -0.0002 |
| | (0.0008) | (0.0009) |
| Number of adults | 0.0326** | -0.0010 |
| | (0.0132) | (0.0142) |
| Self-employed | 0.0789 | 0.0381 |
| | (0.0493) | (0.0659) |
| Employee | 0.0660** | 0.0410 |
| | (0.0322) | (0.0482) |
| Upper secondary education | 0.0351 | 0.0122 |
| | (0.0366) | (0.0498) |
| Tertiary education | 0.0798*** | -0.0444* |
| | (0.0246) | (0.0251) |
| Income: €1,751-€2,900 | 0.0429 | -0.0896** |
| | (0.0394) | (0.0448) |
| Income: €2,900 or more | 0.0262 | -0.0278 |
| | (0.0212) | (0.0251) |
| Rural area/small municipality | 0.1080*** | -0.0581** |
| | (0.0323) | (0.0286) |
| Correlation coefficient $\rho$ | 0.1338** | |
| | (0.0583) | |
| McFadden pseudo $R^2$ | 0.1035 | |
| N | 1,249 | |

**Notes**: The Table reports average marginal effects on the marginal probabilities of using independent and non-independent professional human advice, estimated from a bivariate probit model. Robust standard errors, clustered by macro area, age class, and municipality size, are reported in parentheses below the estimates. The model includes macro area fixed effects.

***, ** and * denote significance at the 1, 5 and 10% levels, respectively

Table 7 – The intensity of robo-advice usage and the demand for professional human advice

| Model: | Bivariate probit | |
|---|---|---|
| Outcome: | Marginal probabilities | |
| | Independent human advice | Non-independent human advice |
| | (1) | (2) |
| Robo-advice: Sometimes | 0.0925** | -0.0066 |
| | (0.0381) | (0.0295) |
| Robo-advice: Often | 0.1079** | -0.0979** |
| | (0.0421) | (0.0449) |
| Robo-advice: Very often | 0.1870** | -0.1112* |
| | (0.0816) | (0.0644) |
| Other control variables | Yes | |
| Correlation coefficient $\rho$ | 0.1355** | |
| | (0.0575) | |
| McFadden pseudo $R^2$ | 0.1096 | |
| N | 1,249 | |

**Notes**: The Table reports average marginal effects on the joint probabilities of using different combinations of professional human advice, estimated from a bivariate probit model. Robust standard errors, clustered macro area, age class, and municipality size, are reported in parentheses below the estimates. The model includes the same control variables as in Table 5 and macro area fixed effects. Complete results are available upon request.
***, ** and * denote significance at the 1, 5 and 10% levels, respectively

**Figures**

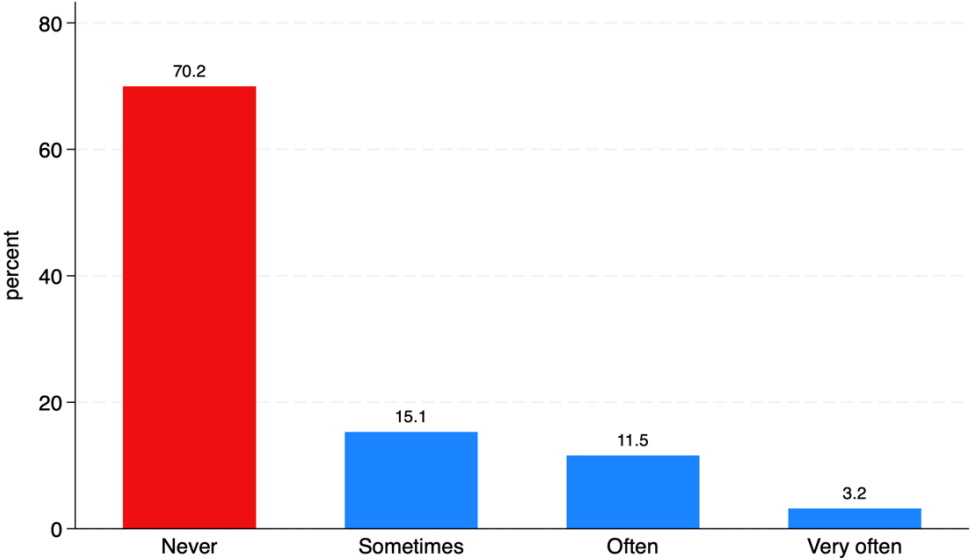

Figure 1 – The intensity of robo-advice usage

# Appendix

Table A1 – Variable definitions

| Variable | Definition |
| --- | --- |
| Robo-advice | Equals 1 if the respondent consulted an online platform for automated financial advice during the last 12 months; 0 otherwise |
| Independent human advice | Equals 1 if the respondent reports that recommendations from an independent fee-based financial advisor significantly influenced the decision to buy a financial product during the last 24 months; 0 otherwise. |
| Non-independent human advice | Equals 1 if the respondent reports that information provided (in person, online or over the phone) by the staff of the financial product provider significantly influenced the decision to buy a financial product during the last 24 months; 0 otherwise |
| Financial literacy | Number of correct answers to the seven financial knowledge questions of the IACOFI questionnaire |
| Subjective financial literacy | Self-rated financial literacy, ranging from 1 ("very low") and 5 ("very high") |
| Digital financial literacy | Number of correct answers to the three digital financial knowledge questions of the IACOFI questionnaire |
| Financial well-being | Financial well-being index defined as the sum of four dummies (then normalized between 0 and 10) indicating whether the respondent is able to covers living expenses with income, can cover living expenses for at least three months in case of loss of the main income source, is able to cope with a major expense without borrowing money, and has money left at the end of the month. |
| Active saver | Equals 1 if the respondent personally saved money (e.g., saving cash at home, paying money into a saving account, saving in an informal savings club, buying bonds, investing in risky financial assets, or saving in other ways other than a pension) during the last 12 months; 0 otherwise |
| Risk propense | Equals 1 if the respondent is willing to take risks at all in taking financial decisions; 0 otherwise |
| Trust in FinTech | Equals 1 if the respondent trusts the financial services provided by online banks and FinTechs; 0 otherwise |
| Digital payments | Score (normalized between 0 and 10) measuring the intensity of a respondent's use of digital payments and online money management instruments during the last 12 months |
| Financial fraud | Equals 1 if the respondent, during the last 24 months, accepted advice to invest in a financial product that was later found to be a scam; 0 otherwise |
| Sustainable investor | Equals 1 if the respondent reports that it is important to invest in companies that strive to minimise their negative impact on the environment, to improve their social impact, and to improve their risk management, ethics and accountability"; 0 otherwise |
| Female | Equals 1 if the respondent is a woman; 0 otherwise |
| Age | Age of the respondent (in years) |
| Number of adults | Number of adults members in the respondent's household |
| Self-employed | Equals 1 if the respondent is self-employed; 0 otherwise |
| Employee | Equals 1 if the respondent is an employee; 0 otherwise |
| Upper secondary education | Equals 1 if the respondent has an upper secondary education; 0 otherwise |
| Tertiary education | Equals 1 if the respondent has a tertiary education; 0 otherwise |
| Income: €1,751-€2,900 | Equals 1 if the respondent's net monthly household disposable income is between 1,751 and 2,900 euro; 0 otherwise |
| Income: €2,900 or more | Equals 1 if the respondent's net monthly household disposable income is above 2,900 euro; 0 otherwise |
| Rural area/small municipality | Equals 1 if the respondent lives in a rural area or in a small municipality; 0 otherwise |

Table A2 – The intensity of robo-advisory usage: addressing potential endogeneity bias

| Model: | Ordered Probit | | | |
|---|---|---|---|---|
| Dependent variable: | Robo-advice intensity | | | |
| Outcome: | Never | Sometimes | Often | Very often |
| | (1) | (2) | (3) | (4) |
| Financial literacy | 0.1262*** | -0.0201*** | -0.0372*** | -0.0689*** |
| | (0.0091) | (0.0015) | (0.0021) | (0.0116) |
| Subjective financial literacy | -0.0260*** | 0.0041*** | 0.0077*** | 0.0142*** |
| | (0.0063) | (0.0012) | (0.0021) | (0.0036) |
| Digital financial literacy | -0.0467*** | 0.0074*** | 0.0138*** | 0.0255*** |
| | (0.0062) | (0.0010) | (0.0016) | (0.0053) |
| Other control variables | Yes | | | |
| Overidentification test | [0.1890] | | | |
| Weak-instruments test: | | | | |
| *Financial literacy* | [0.0000] | | | |
| *Subjective financial literacy* | [0.0000] | | | |
| *Digital financial literacy* | [0.0000] | | | |
| Exogeneity test: | | | | |
| *Financial literacy* | [0.0000] | | | |
| *Subjective financial literacy* | [0.2740] | | | |
| *Digital financial literacy* | [0.7770] | | | |
| N | 4,391 | | | |

**Notes**: The Table reports average marginal effects on the probabilities of the different intensities of robo-advice usage, estimated from an ordered probit model extended to account for the potential endogeneity of financial literacy, subjective financial literacy, and digital financial literacy. Robust standard errors, clustered by macro area, age class, and municipality size, are reported in parentheses below the estimates. All the models include the same control variables as in Table 3 and macro area fixed effects. Complete results are available upon request. The additional instrumental variables for financial literacy are the average financial literacy that individuals of the same gender, living in the same area and belonging to the same age class as the respondent had in 2020 (derived from the 2020 release of the Bank of Italy's IACOFI survey) and a dummy indicating that the respondent is familiar with mortgage loans. The p-values of the overidentification, weak instruments, and exogeneity tests are reported in square brackets.

\*\*\*, \*\* and \* denote significance at the 1, 5 and 10% levels, respectively

Table A3 – Alternative definitions of financial literacy: addressing potential endogeneity bias

| Model: | Probit | Probit | Probit | Probit |
|---|---|---|---|---|
| Dependent variable: | Robo-advice | Robo-advice | Robo-advice | Robo-advice |
| Definition of FL: | IRT | Factor | Big Three | Eight questions |
|  | (1) | (2) | (3) | (4) |
| Financial literacy | -0.1218*** | -0.1115*** | -0.2417*** | -0.1031*** |
|  | (0.0103) | (0.0090) | (0.0170) | (0.0099) |
| Subjective financial literacy | 0.0283*** | 0.0278*** | 0.0292*** | 0.0312*** |
|  | (0.0064) | (0.0064) | (0.0060) | (0.0068) |
| Digital financial literacy | 0.0429*** | 0.0277*** | 0.0441*** | 0.0467*** |
|  | (0.0063) | (0.0042) | (0.0056) | (0.0068) |
| Other control variables | Yes | Yes | Yes | Yes |
| Overidentification test | [0.2610] | [0.3800] | [0.2240] | [0.6680] |
| Weak-instruments test: |  |  |  |  |
| *Financial literacy* | [0.0000] | [0.0000] | [0.0000] | [0.0000] |
| *Subjective financial literacy* | [0.0000] | [0.0000] | [0.0000] | [0.0000] |
| *Digital financial literacy* | [0.0000] | [0.0000] | [0.0000] | [0.0000] |
| Exogeneity test: |  |  |  |  |
| *Financial literacy* | [0.0000] | [0.0000] | [0.0000] | [0.0000] |
| *Subjective financial literacy* | [0.1770] | [0.2430] | [0.2660] | [0.3380] |
| *Digital financial literacy* | [0.6710] | [0.6710] | [0.7060] | [0.8320] |
| N | 4,391 | 4,391 | 4,391 | 4,391 |

**Notes**: The Table reports average marginal effects on the probability of using robo-advisory services, estimated from probit models extended to account for the potential endogeneity of financial literacy, subjective financial literacy, and digital financial literacy and considering alternative definitions of financial literacy. Robust standard errors, clustered by macro area, age class, and municipality size, are reported in parentheses below the estimates. All the models include the same control variables as in Table 2 and macro area fixed effects. Complete results are available upon request. The additional instrumental variables for financial literacy are the average financial literacy that individuals of the same gender, living in the same area and belonging to the same age class as the respondent had in 2020 (derived from the 2020 Bank of Italy's IACOFI survey) and a dummy indicating that the respondent is familiar with mortgage loans. The p-values of the overidentification, weak instruments, and exogeneity tests are reported in square brackets.
***, ** and * denote significance at the 1, 5 and 10% levels, respectively

Table A4 – Robo-advice usage and the demand for alternative combinations of human advice

| | | Bivariate probit | | |
|---|---|---|---|---|
| Model: | | | | |
| Outcome: | | Joint probabilities | | |
| | No human advice (3) | Independent human advice only (4) | Non-independent human advice only (5) | Both types of human advice (6) |
|---|---|---|---|---|
| Robo-advice | -0.0141 | 0.0722*** | -0.0939*** | 0.0358** |
| | (0.0265) | (0.0194) | (0.0264) | (0.0168) |
| Financial literacy | -0.0279*** | -0.0030 | 0.0200*** | 0.0109*** |
| | (0.0070) | (0.0027) | (0.0058) | (0.0033) |
| Subjective financial literacy | -0.0313** | 0.0117 | 0.0002 | 0.0194** |
| | (0.0158) | (0.0074) | (0.0136) | (0.0084) |
| Digital financial literacy | 0.0046 | 0.0141** | -0.0234* | 0.0047 |
| | (0.0164) | (0.0070) | (0.0130) | (0.0083) |
| Financial well-being | -0.0103*** | -0.0003 | 0.0062 | 0.0044** |
| | (0.0037) | (0.0027) | (0.0044) | (0.0022) |
| Active saver | -0.1046*** | 0.0216 | 0.0348 | 0.0482*** |
| | (0.0398) | (0.0148) | (0.0304) | (0.0149) |
| Risk propense | -0.0125 | 0.0456* | -0.0565 | 0.0234 |
| | (0.0329) | (0.0246) | (0.0356) | (0.0181) |
| Digital payments | -0.0205*** | -0.0034 | 0.0164*** | 0.0074*** |
| | (0.0050) | (0.0028) | (0.0057) | (0.0024) |
| Financial fraud | -0.1172* | 0.0744** | -0.0515 | 0.0943** |
| | (0.0606) | (0.0358) | (0.0490) | (0.0449) |
| Sustainable investor | 0.0256 | -0.0150 | 0.0078 | -0.0184 |
| | (0.0244) | (0.0130) | (0.0251) | (0.0118) |
| Female | 0.0537** | 0.0121 | -0.0480** | -0.0178* |
| | (0.0222) | (0.0101) | (0.0195) | (0.0102) |
| Age | -0.0016* | 0.0018*** | -0.0016** | 0.0014*** |
| | (0.0009) | (0.0005) | (0.0008) | (0.0004) |
| Number of adults | -0.0162 | 0.0172** | -0.0164 | 0.0154** |
| | (0.0131) | (0.0077) | (0.0127) | (0.0069) |
| Self-employed | -0.0688 | 0.0307 | -0.0101 | 0.0482 |
| | (0.0596) | (0.0260) | (0.0456) | (0.0355) |
| Employee | -0.0666 | 0.0255 | 0.0006 | 0.0404** |
| | (0.0433) | (0.0188) | (0.0390) | (0.0195) |
| Upper secondary education | -0.0276 | 0.0154 | -0.0075 | 0.0197 |
| | (0.0469) | (0.0195) | (0.0367) | (0.0240) |
| Tertiary education | -0.0087 | 0.0531*** | -0.0710*** | 0.0267* |
| | (0.0254) | (0.0138) | (0.0199) | (0.0139) |
| Income: €1,751-€2,900 | 0.0431 | 0.0466 | -0.0860** | -0.0036 |
| | (0.0316) | (0.0309) | (0.0408) | (0.0160) |
| Income: €2,900 or more | 0.0075 | 0.0203* | -0.0338 | 0.0059 |
| | (0.0230) | (0.0123) | (0.0207) | (0.0120) |
| Rural area/small municipality | -0.0141 | 0.0722*** | -0.0939*** | 0.0358** |
| | (0.0265) | (0.0194) | (0.0264) | (0.0168) |
| N | 1,249 | | | |

**Notes**: The Table reports average marginal effects on the joint probabilities of using different combinations of professional human advice, estimated from a bivariate probit model. Robust standard errors, clustered by macro area, age class, and municipality size, are reported in parentheses below the estimates. The model include macro area fixed effects.
\*\*\*, \*\* and \* denote significance at the 1, 5 and 10% levels, respectively

Table A5 – The demand for professional human advice: addressing potential selectivity bias

| Model: | Bivariate probit with selection | |
|---|---|---|
| Outcome: | Marginal probabilities | |
| | Independent human advice | Non-independent human advice |
| | (1) | (2) |
| Robo-advice | 0.1102*** | -0.0570** |
| | (0.0331) | (0.0283) |
| Other control variables | Yes | |
| Correlation coefficient $\rho$ | 0.1363** | |
| | (0.0645) | |
| Test of no selection bias: | | |
| *Non-independent human advice* | [0.7830] | |
| *Independent human advice* | [0.2170] | |
| N | 4,391 | |

Notes: The Table reports average marginal effects on the marginal probabilities of using independent and non-independent professional human advice, estimated from a bivariate probit model accounting for endogenous sample selection. Robust standard errors, clustered by macro area, age class, and municipality size, are reported in parentheses below the estimates. The model includes the same control variables as in Table 5 and macro area fixed effects. Complete results are available upon request. To improve model identifiability, we include a variable measuring to what extent the respondent makes considered purchases and a dummy indicating that the respondent actively keeps track of money solely in the selection equation. The p-values of the no-selection bias tests are reported in square brackets.
***, ** and * denote significance at the 1, 5 and 10% levels, respectively

Table A6 – The demand for professional human advice: addressing potential endogeneity concerns

| Model: | Endogenous bivariate probit | | Endogenous bivariate probit | |
|---|---|---|---|---|
| Outcome: | Marginal probabilities | | Marginal probabilities | |
| | Independent human advice | Non-independent human advice | Independent human advice | Non-independent human advice |
| | (1) | (2) | (3) | (4) |
| Robo-advice | 0.1678*** | -0.1433*** | 0.1827*** | -0.1262*** |
| | (0.0335) | (0.0260) | (0.0339) | (0.0344) |
| Financial literacy | 0.0104* | 0.0263*** | 0.0125* | 0.0213*** |
| | (0.0053) | (0.0085) | (0.0071) | (0.0083) |
| Other control variables | Yes | | Yes | |
| Correlation coefficient $\rho$ | 0.1258** | | 0.1222** | |
| | (0.0617) | | (0.0548) | |
| Overidentification test | | | [0.2469] | |
| Weak-instruments test: | | | | |
| *Robo-advice* | [0.0000] | | [0.0000] | |
| *Financial literacy* | | | [0.0000] | |
| Exogeneity test: | | | | |
| *Robo-advice* | | | | |
| Non-independent human advice | [0.3640] | | [0.3040] | |
| Independent human advice | [0.4970] | | [0.5040] | |
| *Financial literacy* | | | | |
| Non-independent human advice | | | [0.0920] | |
| Independent human advice | | | [0.7670] | |
| N | 1,249 | | 1,249 | |

Notes: The Table reports average marginal effects on the probabilities of using independent and non-independent professional human advice, estimated from bivariate probit models accounting for the endogeneity of robo-advice usage (columns 1 and 2) and both robo-advice usage and financial literacy (columns 3 and 4). Robust standard errors, clustered by macro area, age class, and municipality size, are reported in parentheses below the estimates. All the models include the same control variables as in Table 5 and macro area fixed effects. Complete results are available upon request. We consider as an additional instrument for robo-advice usage the dummy variable *Trust in FinTech*. The additional instrumental variables for financial literacy are the average financial literacy that individuals of the same gender, living in the same area and belonging to the same age class as the respondent had in 2020 (derived from the 2020 release of the Bank of Italy's IACOFI survey) and a dummy indicating that the respondent is familiar with mortgage loans. The p-values of the overidentification, weak instruments, and exogeneity tests are reported in square brackets.

***, ** and * denote significance at the 1, 5 and 10% levels, respectively